\def\acap{\\ \nonumber \\}
\def\hn{{\hat{r^{'}}}}
\def\nnb{\nonumber }

\def\sif{\sin f}
\def\cif{\cos f}
\def\cu{\cos\left(f + \omega\right)}
\def\su{\sin\left(f + \omega\right)}

\def\kx{{{\hat{k}}_x^{'}}}
\def\ky{{{\hat{k}}_y^{'}}}
\def\kz{{{\hat{k}}_z^{'}}}
\def\Pb{P_{\rm b}}
\def\PbM{P^{'}_{\rm b}}
\def\nk{n_{\rm b}}
\def\nkM{\nk^{'}}
\def\aM{a^{'}}
\def\eM{e^{'}}
\def\eeM{{\eM}^2}
\def\OM{\Om^{'}}
\def\oM{\omega^{'}}
\def\IM{I^{'}}
\def\sIM{\sin\IM}
\def\cIM{\cos\IM}
\def\sIIM{\sin 2\IM}
\def\cIIM{\cos 2\IM}
\def\sIIIM{\sin 3\IM}
\def\cIIIM{\cos 3\IM}
\def\sIMq{\sin^2\IM}

\def\sOM{\sin\OM}
\def\cOM{\cos\OM}
\def\sOOM{\sin 2\OM}
\def\cOOM{\cos 2\OM}
\def\sOOOM{\sin 3\OM}
\def\cOOOM{\cos 3\OM}
\def\sOMq{\sin^2\OM}
\def\cOMq{\cos^2\OM}

\def\sooM{\sin 2\oM}
\def\cooM{\cos 2\oM}

\def\rfr#1{eq. (\ref{#1})}

\def\dert#1#2{\frac{{{d}}{#1}}{{{d}}{#2}}}

\def\virg#1{``#1''}

\def\eqi{\begin{equation}}
\def\eqf{\end{equation}}
\def\eqia{\begin{eqnarray}}
\def\eqfa{\end{eqnarray}}
\def\Om{\mathit{\Omega}}
\def\rp#1#2{{#1\over#2}}
\def\lb#1{\label{#1}}
\def\kap{\bds{\hat{k^{'}}}}

\def\bds#1{\boldsymbol{#1}}

\def\co{\cos\omega}
\def\so{\sin\omega}
\def\coo{\cos 2\omega}
\def\soo{\sin 2\omega}
\def\cO{\cos\Om}
\def\sO{\sin\Om}
\def\cOO{\cos 2\Om}
\def\sOO{\sin 2\Om}

\def\cI{\cos I}
\def\sI{\sin I}
\def\cII{\cos 2I}
\def\sII{\sin 2I}

\def\ee{e^2}

\def\ton#1{\left(#1\right)}
\def\qua#1{\left[#1\right]}
\def\grf#1{\left\{#1\right\}}
\def\ang#1{\left\langle #1\right\rangle}

\documentclass{frontiersFPHY}

\usepackage{amsmath,amsthm,amscd,amssymb,url,lineno,latexsym,wasysym}

\copyrightyear{}
\pubyear{}

\def\journal{Astronomy and Space Sciences}
\def\DOI{10.3389/fspas.2014.00003}
\def\articleType{Research Article}
\def\keyFont{\fontsize{8}{11}\helveticabold }
\def\firstAuthorLast{L Iorio} 
\def\Authors{Lorenzo Iorio\,$^{1,*}$}

\def\Address{$^{1}$Ministero dell'Istruzione, dell'Universit\`{a} e della Ricerca, Istruzione, Bari (BA), Italy }

\def\corrAuthor{\Authors}
\def\corrAddress{Ministero dell'Istruzione, dell'Universit\`{a} e della Ricerca, Istruzione, Viale Unit\`{a} di Italia 68, Bari (BA), 70125, Italy}
\def\corrEmail{lorenzo.iorio@libero.it}

\allowdisplaybreaks[1]

\begin{document}
\onecolumn
\firstpage{1}

\title[Post-Newtonian orbital tidal effects]{Orbital motions as gradiometers for post-Newtonian tidal effects}
\author[L. Iorio]{\Authors}
\address{}
\correspondance{}
\extraAuth{}

\topic{}

\maketitle

\begin{abstract}
The direct long-term changes occurring in the orbital dynamics of a local gravitationally bound binary system $S$ due to the post-Newtonian tidal acceleration caused by an external massive source  are investigated. A class of systems made of a test particle $m$ rapidly orbiting with orbital frequency $\nk$ an astronomical body of mass $M$  which, in turn, slowly revolves around a distant object of mass $M^{'}$ with orbital frequency $\nkM\ll\nk$  is considered. The characteristic frequencies of the non-Keplerian orbital variations of $m$ and of $M$ itself are assumed to be negligible with respect to both $\nk$ and $\nkM$. General expressions for the resulting Newtonian and post-Newtonian tidal orbital shifts of $m$ are obtained. The future missions BepiColombo and JUICE to Mercury and Ganymede, respectively, are considered in view of a possible detection. The largest effects, of the order of  $\approx 0.1-0.5$ milliarcseconds per year (mas yr$^{-1}$), occur for the Ganymede orbiter of the JUICE mission. Although future improvements in spacecraft tracking and orbit determination might, perhaps, reach the required sensitivity, the systematic bias represented by the other known orbital perturbations of both Newtonian and post-Newtonian origin would be overwhelming. The realization of a dedicated artificial mini-planetary system to be carried onboard and Earth-orbiting spacecraft is considered as well. Post-Newtonian tidal precessions as large as $\approx 1-10^2$ mas yr$^{-1}$ could be obtained, but the quite larger Newtonian tidal effects would be a major source of systematic bias because of the present-day percent uncertainty in the product of the Earth's mass times the Newtonian gravitational parameter.

\tiny
 \keyFont{ \section{Keywords:} General relativity and gravitation; Experimental studies of gravity; Experimental tests of gravitational theories; Celestial mechanics; Spacecraft }
\end{abstract}


\section{Introduction}\lb{intro}
Gravitation is one of the known fundamental interactions of physics, and the General Theory of Relativity (GTR) is, at present, its best theoretical description  \cite{2009SSRv..148....3W}. As such, GTR is one of the pillars of our knowledge of Nature; intense experimental and observational scrutiny is required not only to gain an ever-increasing confidence about it, but also to explore the borders of the realm of its validity at different scales. To this aim, a variety of different theoretical, experimental and observational approaches are required to extend the frontiers of our knowledge of the gravitational phenomena.  Are there some founded hopes to testing newly predicted gravitational effects in the near future in some suitable astronomical and astrophysical laboratories?  What are the possibilities opened up by forthcoming space-based missions?  The present paper will try to address these questions by looking at certain effects that the components of the Riemann spacetime curvature tensor are expected to induce on local systems according to GTR.

The internal dynamics of a gravitationally bound  binary system immersed in the external gravitational field of a massive rotating body is tidally affected at both the Newtonian and the post-Newtonian level \cite{1977ApJ...216..591M, 2002CQGra..19.4231C, 2006CQGra..23.4021C, 2011rcms.book.....K}.
In this paper, we will look in detail at some of the post-Newtonian orbital effects of tidal origin arising in the relative motion of a restricted two-body system, and at the possibility of detecting them in either natural or artificial space-based scenarios. Special cases widely treated in the literature are the post-Newtonian tidal effects of the rotating Sun's field in the Earth-Moon system \cite{Bra80, 1982PhRvL..49.1542M, 1986PhLA..115..333M, 1989PhRvD..39.2441G, 1991NCimB.106..545M, 2001LNP...562..310M} and of the spinning Earth itself in arrays of spaceborne artificial gradiometers \cite{1982PhRvL..49.1542M, 1985PhLA..109...19T, 1989PhRvD..39.2825M, 1989AdSpR...9...41P, 1992LNP...410..131T, 2008GReGr..40..907P, 2014GReGR..46....1L}. Our calculation will have a broad range of validity. Indeed, while, on the one hand, certain  assumptions on the characteristic orbital frequencies of the three-body system considered will be necessarily  made, on the other hand, we will remove the limitations existing in the literature \cite{2006CQGra..23.4021C} on either the orientation of the spin axes of the external objects and on the orbital configurations of the moving bodies.

The paper is organized as follows. In Section \ref{rates_m}, the long-term rates of change of the orbital parameters of the test particle of the restricted two-body system are calculated by keeping the  elements of a generic tidal matrix constant over the orbital period of the particle around its primary. In Section \ref{rates_M}, the direct orbital effects due to both the gravitoelectric and the gravitomagnetic tidal matrices are obtained by averaging their elements over the orbital period of the motion around the distant body.
Section \ref{expe} is devoted to exploring some experimental possibilities offered by forthcoming spacecraft-based missions to astronomical bodies.
Section \ref{concludi} summarizes our findings.
\section{The long-term orbital rates of change averaged over $P_{\rm b}$}\lb{rates_m}
Let us consider an isolated rotating body of mass $M^{'}$, equatorial radius $R^{'}$ and proper angular momentum ${\bds{J}}^{'}$ at rest in some parameterized post-Newtonian coordinate system ${\mathcal{K}}^{'}$ whose spatial axes point to distant stars; as such, ${\mathcal{K}}^{'}$ is kinematically and dynamically non-rotating \cite{1989NCimB.103...63B}. Let a local gravitationally bound system $S$ move geodesically around $M^{'}$; $S$ is assumed to be made of a body of mass $M$, equatorial radius $R$ and proper angular momentum $\bds J$,  and of a test particle of mass $m$ revolving about $M$ itself. For the sake of simplicity, we will assume $ m\ll M \ll M^{'}$; examples of such scenarios are spacecraft like MESSENGER \cite{2007SSRv..131....3S} and the forthcoming BepiColombo \cite{2010P&SS...58....2B} to orbit Mercury in the field of the Sun as well as the future JUICE mission \cite{2013P&SS...78....1G} to orbit Ganymede and to study the Jovian system.
To the Newtonian level, the spatial trajectory of $S$ about $M^{'}$ can be parameterized in terms of the usual Keplerian orbital elements. They are the semimajor axis  $\aM$, the eccentricity $\eM$, the inclination $\IM$ to the reference $\grf{x^{'},y^{'}}$ plane of ${\mathcal{K}}^{'}$, the longitude of the ascending node $\OM$, and the argument of pericenter $\oM$;  $\nkM=\sqrt{GM^{'} {\aM}^{-3}}$ is the Keplerian orbital frequency, where $G$ is the Newtonian constant of gravitation. In general, such orbital parameters do not stay constant because of the well known Newtonian and post-Newtonian departures from spherical symmetry of the field of $M^{'}$. They induce the post-Newtonian Einstein (gravitoelectric) \cite{1915SPAW...47..831E, 1986Natur.320...39N} and Lense-Thirring (gravitomagnetic) \cite{LT18} orbital precessions as well as the classical ones due to the oblateness $J_2^{'}$ of $M^{'}$ and, possibly, to its other multipoles of higher order \cite{BeFa2003}.

Let a local inertial frame $\mathcal{K}$, attached to $M$, be parallel transported along the geodesic worldline of $M$ through the spacetime of $M^{'}$ \cite{Fermi22, Levi27, 1927RSPTA.226...31S}. As such, the spatial axes of $\mathcal{K}$ change naturally their orientation with respect to the fixed \virg{Copernican} spatial axes of ${\mathcal{K}}^{'}$ because of the geodesic motion of $\mathcal{K}$ itself through the external spacetime deformed by $M^{'}$ and ${\bds{J}}^{'}$. As such, $\mathcal{K}$ is also said to be  kinematically rotating because of the resulting de Sitter-Fokker (gravitoelectric) \cite{deSitter1916, Fokker1921} and Pugh-Schiff (gravitomagnetic) precessions\footnote{A comoving coordinate system is said to be kinematically nonrotating if it is corrected for the post-Newtonian precessions of its axes.}  of its axes with respect to those of  ${\mathcal{K}}^{'}$, but it is dynamically nonrotating because of the absence of Coriolis and centrifugal inertial forces \cite{1989NCimB.103...63B}.
To the Newtonian level, also the motion of $m$ about $M$ will be parameterized in terms of a set of Keplerian orbital elements $a,e,I,\Om,\omega$ in such a way that $\nk=\sqrt{GM a^{-3}}$ denotes its Keplerian orbital frequency.
Usually, the frequencies of the de Sitter-Fokker and Pugh-Schiff precessions are quite smaller than both  $\nkM$ and $\nk$; for a critical discussion tidal phenomena occurring in the Sun-Earth-Moon system over timescales comparable to or larger than the de Sitter-Fokker and Pugh-Schiff ones, see \cite{1989PhRvD..39.2441G, 1991NCimB.106..545M}.
Thus, we can safely assume $\mathcal{K}$ as kinematically nonrotating over timescales comparable to the orbital periods $\Pb=2\pi\nk^{-1}$ and $\PbM=2\pi{\nkM}^{-1}$ of the three-body system considered.
As a further assumption, we will consider the local motion of $m$ about $M$ much faster than the one of $S$ itself around $M^{'}$, i.e.
$\nkM\ll\nk.$ In general, the internal dynamics of $S$ is  not purely Keplerian because of possible departures from sphericity of $M$ and of the post-Newtonian components of the field of $M$. As such, the orbit of $m$ with respect to $M$ undergoes the well-known Newtonian and post-Newtonian orbital precessions. As it occurs in the  systems considered here, the timescales of such changes are quite longer than the orbital periods $\Pb,\PbM$, i.e. $\dot\Psi\ll\nkM$, where $\Psi$ denotes a generic precessing osculating Keplerian orbital element of $m$.

At both the Newtonian and the post-Newtonian level, the internal dynamics of $S$ is locally affected also by tidal effects due to its motion through the external deformed spacetime of $M^{'}$.

The tidal acceleration experienced by $m$  is of the form \cite{1989PhRvD..39.2825M}
\eqi{\bds A}_{\rm tid} = -\textbf{\textsf{K}}\bds r\lb{Atidal},\eqf
where the elements of the tidal matrix \textbf{\textsf{K}}
\eqi \mathrm{K}_{ij} = {\mathcal{R}}_{0i0j},\ i,j=1,2,3\eqf are the tetrad components of the curvature Riemann tensor evaluated onto the geodesic of the observer in $\mathcal{K}$, and have dimensions of T$^{-2}$.
It is
\eqi \textbf{\textsf{K}} = {\textbf{\textsf{K}}}^{(\rm N)} + {\textbf{\textsf{K}}}^{(\rm GE)} + {\textbf{\textsf{K}}}^{(\rm GM)},\eqf
with the Newtonian (N), gravitoelectric (GE) and gravitomagnetic (GM) tidal matrices given by \cite{1989PhRvD..39.2825M}
\begin{align}
\mathrm{K}_{ij}^{(\rm N)} \lb{KN} & =\rp{GM^{'}}{{r^{'}}^3}\ton{\delta_{ij} - 3\hn_i\hn_j}, \\ \nonumber \\
\mathrm{K}_{ij}^{(\rm GE)} \nonumber \lb{KGE} & = -\rp{G^2 {M^{'}}^2}{c^2{r^{'}}^4 }\ton{3\delta_{ij} -9\hn_i\hn_j  } +  \\ \nonumber \\
& + \rp{GM^{'}}{c^2{r^{'}}^3 }\grf{3\qua{{v^{'}}^2\delta_{ij}-{v^{'}}_i {v^{'}}_j + 3 \ton{\bds {v^{'}}\bds\cdot \bds{\hat{{r^{'}}}}}\hn_{(i}{v^{'}}_{j)} } -3\ton{\bds {v^{'}}\bds\cdot\bds{\hat{{r^{'}}}}}^2\delta_{ij} -6\hn_i\hn_j {v^{'}}^2  }, \\ \nonumber \\
\mathrm{K}_{ij}^{(\rm GM)} \lb{KGM} \nonumber & = -\rp{6GJ^{'}}{c^2{r^{'}}^4 }\grf{3\ton{\bds {v^{'}}\bds\times\kap}_{(i}\hn_{j)} + \ton{\bds{\hat{{r^{'}}}}\bds\times\kap}_{(i}{v^{'}}_{j)}
+ \bds{\hat{{r^{'}}}}\bds\cdot\ton{\bds {v^{'}}\bds\times \kap}\ton{\delta_{ij}-5\hn_i\hn_j }  -\right.\\ \nonumber \\
& -\left.  5\ton{\bds{\hat{{r^{'}}}}\bds\cdot\bds {v^{'}} }\ton{\bds{\hat{{r^{'}}}}\bds\times\kap }_{(i}\hn_{j)} }.
\end{align}
In \rfr{KN}-\rfr{KGM}, which are symmetric and traceless, $c$ is the speed of light in vacuum, $\bds{\hat{r^{'}}}={\bds {r^{'}}}/r^{'}$ is the versor of the position vector $\bds{r^{'}}$ from $M^{'}$ to $M$, $\bds{v^{'}}$ is the velocity vector of $M$ with respect to $M^{'}$, $\delta_{ij}$ is the Kronecker symbol,  $\kap$ is the unit vector of the spin axis of $M^{'}$, the symbols $\bds\cdot$ and $\bds\times$ denote the usual scalar and cross products among vectors, and parentheses around indices denote symmetrization.

The tidal acceleration of \rfr{Atidal} can be considered as a small perturbation ${\bds A}_{\rm pert}$ of the Newtonian monopole of $M$. As such, its impact on the orbital dynamics of $m$ can be treated perturbatively with standard techniques.
By recalling
the condition $\nkM\ll\nk$, the elements of the tidal matrix \textbf{\textsf{K}} can be considered as constant over an orbital period $P_{\rm b}$. Thus, by evaluating the right-hand-sides of the Gauss equations \cite{1976AmJPh..44..944B}
\begin{align}
\dert a t \lb{Gdadt} & = \rp{2}{\nk\sqrt{1-e^2}}\ton{e A_R \sif + \rp{p}{r} A_T}, \\ \nonumber \\
\dert e t & =  \rp{\sqrt{1-e^2}}{\nk a}\grf{ A_R\sin f + A_T\qua{\cos f +\rp{1}{e} \ton{ 1 - \rp{r}{a} } } }, \\ \nonumber \\
\dert I t & =\rp{\cos\ton{\omega + f}}{\nk a\sqrt{1-e^2}}\ton{\rp{r}{a}}A_N, \\ \nonumber \\
\dert \Om t \lb{GdOdt} & =\rp{\sin\ton{\omega + f}}{\nk a\sI\sqrt{1-e^2}}\ton{\rp{r}{a}}A_N, \\ \nonumber \\
\dert\omega t  +\cI\dert\Om t \lb{Gdodt} & = \rp{\sqrt{1-e^2}}{\nk a e}\qua{- A_R \cos f + A_T \ton{ 1 + \rp{r}{p} } \sin f}.
\end{align}
onto the unperturbed Keplerian ellipse
\eqi r = \rp{p}{1+e\cos f},\eqf where $p=a\left(1-e^2\right)$ is the semilatus rectum and $A_R,A_T,A_N$ are the radial, transverse and out-of-plane components of \rfr{Atidal}, the orbital variations of $m$ averaged over $P_{\rm b}$ can be calculated.
To this aim, let us note that, in principle, the average should be made by means of \cite{1991ercm.book.....B, 2014PhRvD..89d4043W}
\eqi
\dert f t = \nk\ton{\rp{a}{r}}^2\sqrt{1-e^2} -\ton{\dert\omega t +\cI\dert\Om t},\lb{dfdtfull}
\eqf
where $t$ is the proper time along the observer's geodesic, $f$ is the true anomaly, and $\dot\Om,\dot\omega$ are to be intended as the instantaneous, non-averaged precessions of the node and the pericenter \cite{1991ercm.book.....B}. Indeed, $f$ is reckoned from the pericenter position, which, in general, changes because of possible variations of $\Om$ and $\omega$ due to the non-Keplerian Newtonian and post-Newtonian effects within $S$. As such, the instantaneous expressions for $\dot\Om$ and $\dot\omega$  in \rfr{dfdtfull} should  be taken from \rfr{GdOdt}-\rfr{Gdodt} themselves evaluated for the specific Newtonian and post-Newtonian perturbations \cite{1991ercm.book.....B}. However, by limiting ourselves just to the order $\mathcal{O}\ton{c^{-2}}$, and by neglecting small mixed terms of order $\mathcal{O}\ton{J_2 c^{-2}}$ arising from the interplay between the external tidal and the local Newtonian  perturbations due to $J_2$,
the approximate expression \eqi
\dert f t \lb{dfdtKep} = \nk\ton{\rp{a}{r}}^2\sqrt{1-e^2}
\eqf
can be used by integrating over $f$ between $0$ and $2\pi$.
By using the following Keplerian expressions for the position $\bds r$
\begin{align}
x \lb{xKep} & = r\qua{\cO\cu-\cI\sO\su}, \\ \nonumber \\
y &= r\qua{\sO\cu + \cI\cO\su}, \\ \nonumber \\
z &= r\qua{\sI\su},
\end{align}
and the velocity $\bds v$
\begin{align}
v_x &= -\rp{a\nk\grf{\cI\sO\qua{e\co+\cu} +\cO\qua{e\so+\su}  }}{\sqrt{1-\ee}}, \\ \nonumber \\
v_y &=\rp{a\nk\grf{\cI\cO\qua{e\co+\cu} -\sO\qua{e\so+\su}  }  }{\sqrt{1-\ee}}, \\ \nonumber \\
v_z \lb{vzKep} &=\rp{a\nk\sI\qua{e\co+\cu}}{\sqrt{1-\ee}},
\end{align}
it is possible to compute from \rfr{xKep}-\rfr{vzKep}
the unit vector $\bds{\hat{L}}$ along the orbital angular momentum
as
\eqi \bds{\hat{L}} = \rp{\bds r\bds\times\bds v}{\left| \bds r\bds\times\bds v \right|}. \eqf
Then, the radial, transverse and normal components of \rfr{Atidal} turn out to be
\begin{align}
A_R \nonumber \lb{AtidR} & = {\bds A}_{\rm tid}\bds\cdot\bds{\hat{r}}= \rp{\mathrm{K}a\ton{1-\ee}}{1 + e\cos f }\grf{
\left[\cI \su \sO - \cu \cO\right]\cdot\right.\\ \nonumber \\
\nonumber &\cdot\left.  \left[\sif \left(K_{21} \cI \co \cO + \left(K_{22} + K_{33}\right) \so \cO + K_{31} \co \sI\right) +\right.\right.\\ \nonumber \\
\nonumber &+\left.\left.  \cif \left(K_{11} \co \cO + \left(K_{21} \cI \cO + K_{31} \sI\right) \so\right) + \right.\right.\\ \nonumber \\
\nonumber & +\left.\left. \left(K_{21} \cu + \left(K_{22} + K_{33}\right) \cI \su\right) \sO\right] + \right.\\ \nonumber \\
\nonumber & +\left. \left(\cI \cO \su + \cu \sO\right) \left(\cu \left(K_{21} \cO + K_{22} \sO\right) + \right.\right.\\ \nonumber \\
\nonumber & +\left.\left. \su \left(K_{32} \sI + \cI \left(K_{22} \cO - K_{21} \sO\right)\right)\right) + \right.\\ \nonumber \\
\nonumber & +\left. \sI \su \left(\cu \left(K_{31} \cO + K_{32} \sO\right) + \right.\right.\\ \nonumber \\
 & +\left.\left. \su \left(K_{33} \sI + \cI \left(K_{32} \cO - K_{31} \sO\right)\right)\right) }, \\ \nonumber \\
A_T \nonumber & = {\bds A}_{\rm tid}\bds\cdot\ton{ {\bds{\hat{L}}}{\bds\times}{\bds{\hat{r}}} }= -\rp{\mathrm{K}a\ton{1-\ee}}{4\ton{1+e\cos f} }\grf{ \left(-K_{33} + \left(2 K_{22} + K_{33}\right) \cOO - \right.\right.\\ \nonumber \\
\nonumber & -\left.\left. 2 K_{21} \sOO\right) \cos ^2 I \sin \left(2f + 2\omega\right) + 2 \left(2 \sI  \left(K_{32} \cO - K_{31} \sO\right)\sin \left(2f + 2\omega \right) + \right.\right. \\ \nonumber \\
\nonumber & + \left.\left.  \left[2 K_{21} \cOO + \left(2 K_{22} + K_{33}\right) \sOO\right]\right) \cI\cos \left(2f + 2\omega \right) + \right. \\ \nonumber \\
\nonumber & +\left. 4  \sI \left(K_{31} \cO + K_{32} \sO\right)\cos \left(2f + 2\omega \right) +
 \left[\left(2-\cII\right) K_{33}  + \right.\right. \\ \nonumber \\
& +\left.\left. \left(2 K_{22} + K_{33}\right) \cOO - 2 K_{21} \sOO\right]\sin \left(2f + 2\omega \right)}, \\ \nonumber \\
A_N \nonumber \lb{AtidN} & = {\bds A}_{\rm tid}\bds\cdot{\bds{\hat{L}}} = -\rp{\mathrm{K}a\ton{1-\ee}}{4\ton{1+e\cos f} }\grf{4 \cI  \left(K_{31} \cO + K_{32} \sO\right)\cos \left(f + \omega \right) + \right. \\ \nonumber \\
\nonumber & + \left. 2  \sI \left[-2 K_{21}\cOO - \left(2 K_{22} + K_{33}\right) \sOO\right]\cos \left(f + \omega \right) + \right.\\ \nonumber \\
\nonumber &+\left. 4 \cII \left(K_{32} \cO - K_{31}\sO\right) \sin \left(f + \omega\right) + \right.\\ \nonumber \\
& +\left. \sII \left[3 K_{33} - \left(2 K_{22} + K_{33}\right) \cOO + 2 K_{21} \sOO\right]\sin \left(f + \omega\right) },
\end{align}
where K is the dimensional scaling factor of the tidal matrix considered having dimensions of T$^{-2}$, while the dimensionless coefficients $K_{ij},\ i,j=1,2,3$ depend only on the orbital parameters of $M$ and on the orientation of $\kap$.
Thus, by inserting \rfr{AtidR}-\rfr{AtidN} in  the right-hand-sides of \rfr{Gdadt}-\rfr{Gdodt} and averaging them over $P_{\rm b}$ with \rfr{dfdtKep}, one finally has
\begin{align}
\ang{\dert a t}_{\Pb} \lb{dadtm}& = 0, \\ \nonumber \\
\ang{\dert e t}_{\Pb} \lb{dedt} \nonumber & = \rp{5\mathrm{K}e\sqrt{1-e^2}}{8\nk}\grf{-2 \soo \sO \left(2 K_{21} \cO + K_{33}
\sO\right) \cos^2 I +  \right. \\ \nonumber \\
\nonumber &+ \left. 2\left[2 \sI \soo \left(K_{32} \cO - K_{31} \sO\right) + \coo \left(2 K_{21} \cOO + \right.\right.\right.\\ \nonumber \\
\nonumber & +\left.\left.\left. \left(2 K_{22}+ K_{33}\right) \sOO\right)\right] \cI + 4 \coo \sI \left(K_{31} \cO + K_{32} \sO\right) +\right.\\ \nonumber \\
\nonumber & +\left. \soo \left[K_{22} \left(\cII + 3\right) \cOO + K_{33} \left(-\cII + \cOO + 2\right) - \right.\right. \\ \nonumber \\
&-\left.\left. 2 K_{21}\sOO\right]}, \\ \nonumber \\
\ang{\dert I t}_{\Pb} \nonumber \lb{dIdtm} & = -\rp{\mathrm{K}}{8\nk\sqrt{1-e^2}}\grf{10 e^2\cII \soo \left(K_{32} \cO - K_{31} \sO\right) - \right.\\ \nonumber \\
\nonumber &-\left. \frac{5}{2} e^2\sII \soo \left[-3 K_{33} + \left(2 K_{22} + K_{33}\right) \cOO - 2 K_{21} \sOO\right]  + \right.\\ \nonumber \\
\nonumber &+\left. 2 \cI  \left(5 e^2\coo + 3 e^2 + 2 \right) \left(K_{31} \cO + K_{32} \sO\right) -  \right.\\ \nonumber \\
          & -\left. \left(5 e^2\coo + 3 e^2 + 2 \right) \sI \left[2 K_{21}\cOO + \left(2 K_{22} + K_{33}\right) \sOO\right]}, \\ \nonumber \\
\ang{\dert\Om t}_{\Pb} \nonumber & =  -\rp{\mathrm{K}\csc I}{16\nk\sqrt{1-e^2}}\grf{20e^2\soo \left(\cI \left(K_{31} \cO + K_{32} \sO\right) - \right.\right.\\ \nonumber \\
\nonumber &-\left.\left. \sI \left(K_{21} \cOO + \left(2 K_{22} + K_{33}\right) \cO \sO\right)\right) + \right.\\ \nonumber \\
\nonumber & +\left. 4 \cII  \left(5 e^2\coo - 3 e^2 - 2 \right)\left(K_{31} \sO - K_{32} \cO\right) + \right.\\ \nonumber \\
\nonumber & + \left. \left(5 e^2\coo - 3 e^2 - 2 \right) \sII \left[-3 K_{33} + \left(2 K_{22} + K_{33}\right) \cOO - \right.\right.\\ \nonumber \\
& - \left.\left. 2 K_{21} \sOO\right]},  \\ \nonumber \\
\ang{\dert\omega t}_{\Pb} +\cI\ang{\dert\Om t}_{\Pb}  \lb{dodtm} \nonumber & = \rp{\mathrm{K}\sqrt{1-e^2}}{16\nk}\grf{-40 \sI \soo \left(K_{31} \cO + K_{32} \sO\right) + \right.\\ \nonumber \\
\nonumber & + \left. 4 \left(5 \coo - 3\right) \sII \left(K_{32}\cO - K_{31} \sO\right) - \right.\\ \nonumber \\
\nonumber & - \left. 20 \cI \soo \left[2 K_{21} \cOO + \left(2 K_{22} + K_{33}\right)\sOO\right] + \right.\\ \nonumber \\
\nonumber & + \left. \cII \left(5 \coo - 3\right) \left[-3 K_{33} + \left(2 K_{22} + K_{33}\right)\cOO - 2 K_{21} \sOO\right] + \right.\\ \nonumber \\
          & +\left. 3 \left(5 \coo +1\right) \left[K_{33} + \left(2 K_{22} + K_{33}\right) \cOO - 2 K_{21}\sOO\right]}.
\end{align}
 The long-term rates of \rfr{dadtm}-\rfr{dodtm} are valid for any symmetric and traceless tidal-type perturbation of the form of \rfr{Atidal} whose coefficients can be considered as constant over the characteristic orbital frequency $\nk$ of the local binary system considered.  As such, \rfr{dadtm}-\rfr{dodtm} are not limited just to \rfr{KGE}-\rfr{KGM}. Moreover,  \rfr{dadtm}-\rfr{dodtm} hold for a general orbital configuration of the test particle $m$ since no a priori simplifying assumptions concerning its eccentricity and inclination were made. As shown by \rfr{dedt}, the eccentricity of a circular orbit is not affected by a tidal-type perturbation.
\section{The long-term orbital rates of change averaged over $\PbM$}\lb{rates_M}
In general, \rfr{dadtm}-\rfr{dodtm} may not be regarded as truly secular rates over timescales arbitrarily long  because of the slow time dependence encoded in both the tidal matrix elements themselves and in the orbital elements  of $m$, collectively denoted as $\grf{\Psi}$, due to possible non-Keplerian local effects taking place in the non-spherically symmetric field of $M$.

Let us, now, assume that the characteristic timescales $P_{\Psi}$ of all the non-Keplerian orbital effects within $S$ are much longer than the orbital period $\PbM$ of $S$ itself about $M^{'}$. It is a reasonable tenet, satisfied in several astronomical scenarios of potential experimental interest. Thus, it is possible to perform a further average of \rfr{dadtm}-\rfr{dodtm} over $\PbM$ by keeping $a,e,I,\Om,\omega$ constant over the integration with respect to some fast variable\footnote{It turned out computationally more convenient to adopt the true anomaly $f^{'}$.} of the motion of $M$ around $M^{'}$.
%
%

The direct effects of order $\mathcal{O}\ton{c^{-2}}$ can be obtained by evaluating the post-Newtonian tidal matrices \rfr{KGE}-\rfr{KGM} onto an unchanging Keplerian ellipse as reference unperturbed trajectory.
%
%
%
%

Below, the averaged tidal matrix elements of \rfr{KN}-\rfr{KGM}, computed to order $\mathcal{O}\ton{c^{-2}}$ and to zero order in $J_2^{'}$, are listed. They are to be inserted in \rfr{dadtm}-\rfr{dodtm} to have the direct long-term rates of change of $m$ averaged over $\PbM$.

As far as the Newtonian tidal matrix of \rfr{KN} is concerned, its average,  to the zero order in $J_2^{'}$, is
\begin{align}
\ang{{\mathrm{K}}_{11}^{\rm (N)}}_{\PbM} \lb{KN11} & = -\rp{GM^{'}\ton{ 1 + 3\cIIM + 6\sin^2\IM\cOOM  }}{{8\aM}^3\ton{1-{\eM}^2}^{3/2}},\\ \nonumber \\
\ang{{\mathrm{K}}_{22}^{\rm (N)}}_{\PbM} \lb{KN22} & = -\rp{GM^{'}\ton{ 1 + 3\cIIM - 6\sin^2\IM\cOOM  } }{{8\aM}^3\ton{1-{\eM}^2}^{3/2}},\\ \nonumber \\
\ang{{\mathrm{K}}_{33}^{\rm (N)}}_{\PbM} \lb{KN33} & =\rp{GM^{'}\ton{1 + 3\cIIM}}{{4\aM}^3\ton{1-{\eM}^2}^{3/2}},\\ \nonumber \\
\ang{{\mathrm{K}}_{12}^{\rm (N)}}_{\PbM} \lb{KN12} & = -\rp{3GM^{'}\sin^2\IM\sOOM}{{4\aM}^3\ton{1-{\eM}^2}^{3/2}},\\ \nonumber \\
\ang{{\mathrm{K}}_{13}^{\rm (N)}}_{\PbM}  \lb{KN13} & = \rp{3GM^{'}\sIIM\sOM}{{4\aM}^3\ton{1-{\eM}^2}^{3/2}},\\ \nonumber \\
\ang{{\mathrm{K}}_{23}^{\rm (N)}}_{\PbM} \lb{KN23} & = -\rp{3GM^{'}\sIIM\cOM}{{4\aM}^3\ton{1-{\eM}^2}^{3/2}}.
\end{align}

For the post-Newtonian gravitoelectric tidal field of $M^{'}$, \rfr{KGE} yields
\begin{align}
\ang{{\mathrm{K}}_{11}^{\rm (GE)}}_{\PbM} \nnb \lb{KGE11} & = -\rp{3 G^2 {M^{'}}^2 \eeM}{32 c^2{\aM}^4\ton{1-\eeM}^{5/2}}\grf{12 \cIIM + 2 \cooM - 2\cIIM\cooM + \right.\acap
\nnb & + \left. \cOOM  \left[24 \sIMq + 2 \left(\cIIM + 3\right) \cooM \right] - \right.\acap
 & -\left. 8 \cIM  \sOOM \sooM + 4},\acap
\ang{{\mathrm{K}}_{22}^{\rm (GE)}}_{\PbM} \nnb \lb{KGE22} & = \rp{3 G^2 {M^{'}}^2 \eeM}{32 c^2{\aM}^4\ton{1-\eeM}^{5/2}}\grf{-12 \cIIM - 2 \cooM + 2\cIIM\cooM + \right.\acap
\nnb & + \left. \cOOM  \left[24 \sIMq + 2 \left(\cIIM + 3\right) \cooM \right] - \right.\acap
& -\left. 8 \cIM  \sOOM \sooM - 4},\acap
\ang{{\mathrm{K}}_{33}^{\rm (GE)}}_{\PbM} \lb{KGE33} & = \rp{3 G^2 {M^{'}}^2 \eeM}{4 c^2{\aM}^4\ton{1-\eeM}^{5/2}}\ton{\sIMq\cooM + 3 \cIIM + 1},\acap
\ang{{\mathrm{K}}_{12}^{\rm (GE)}}_{\PbM} \nnb \lb{KGE12} & = -\rp{3 G^2 {M^{'}}^2 \eeM}{16 c^2{\aM}^4\ton{1-\eeM}^{5/2}}\grf{4 \cIM \cOOM \sooM +  \right.\acap
& + \left. \left[12 \sIMq + \left(\cIIM + 3\right) \cooM \right] \sOOM},\acap
\ang{{\mathrm{K}}_{13}^{\rm (GE)}}_{\PbM}  \lb{KGE13} & = -\rp{3 G^2 {M^{'}}^2 \eeM}{4 c^2{\aM}^4\ton{1-\eeM}^{5/2}}\grf{\sIM \left[\cOM \sooM + \cIM \left(\cooM - 6\right) \sOM\right]},\acap
\ang{{\mathrm{K}}_{23}^{\rm (GE)}}_{\PbM}  \lb{KGE23} & = \rp{3 G^2 {M^{'}}^2 \eeM}{4 c^2{\aM}^4\ton{1-\eeM}^{5/2}}\grf{\sIM \left[\cIM \left(\cooM - 6\right)\cOM - \sooM\sOM\right]}.
\end{align}
In the limit of $\eM\rightarrow 0$, \rfr{KGE11}-\rfr{KGE23} vanish.

For the post-Newtonian gravitomagnetic tidal field of $M^{'}$ due to $\bds{{J^{'}}}$, from \rfr{KGM} one obtains
\begin{align}
\ang{{\mathrm{K}}_{11}^{\rm (GM)}}_{\PbM} \nnb \lb{KLT11} & = -\rp{3GJ^{'}\nkM}{64 c^2{\aM}^3\ton{1-\eeM}^3}\grf{
40 \eeM \sooM \left(2 \kz \cIIM + 3 \kx \sIIM \sOM\right) \sOOM  + \right.\acap
\nnb & + \left. 5 \cooM  \left(12 \sIIIM \left(\kx \sOM - \ky \cOM\right) \sOMq + \right.\right. \acap
\nnb & +\left.\left. \sIM \left(\ky \left(\cOM + 15 \cOOOM\right) - 3 \kx \left(\sOM + 5 \sOOOM\right)\right) \right) \eeM  - \right. \acap
\nnb & -\left. 20  \left(3 \eeM  + 2 \right) \kz \cIIIM - 4 \cIM  \left(5 \kz \cooM  \left(6 \sIMq + \right.\right.\right. \acap
\nnb & + \left.\left.\left. \left(3 \cIIM + 1\right) \cOOM \right) \eeM  - 10 \ky \sIM\sooM \left(\sOM - 3 \sOOOM\right) \eeM  +  \right.\right.\acap
\nnb & +\left.\left. \left(3 \eeM  + 2 \right) \kz  \left(20 \cOOM \sIMq + 3 \right) \right) + \right. \acap
\nnb & + \left. 2  \left(3 \eeM  + 2 \right)  \left(20 \sIIIM \left(\ky \cOM - \kx \sOM\right) \sOMq + \right.\right. \acap
& + \left.\left. \sIM \left(\ky \left(\cOM + 15 \cOOOM\right) - 3 \kx \left(\sOM + 5 \sOOOM\right)\right) \right)
},\acap
\ang{{\mathrm{K}}_{22}^{\rm (GM)}}_{\PbM} \nnb  \lb{K22}& = -\rp{3GJ^{'}\nkM}{64 c^2{\aM}^3\ton{1-\eeM}^3}\grf{
20 \eeM\kx \left(\cOM + 3 \cOOOM\right) \sIIM \sooM   - \right.\acap
\nnb & - \left. 80\eeM \kz \cIIM \sooM \sOOM   - 20  \left(3 \eeM  + 2 \right) \kz \cIIIM + \right.\acap
\nnb & + \left. 20  \left(3\eeM \cooM   - 6 \eeM  - 4 \right) \cOMq \sIIIM \left(\kx \sOM - \ky \cOM\right) + \right. \acap
\nnb & +\left. 4 \eeM\cIM  \left(5 \kz \cooM  \left(\left(3 \cIIM + 1\right) \cOOM - 6 \sIMq \right)   + \right.\right.\acap
\nnb & +\left.\left. 120\eeM \ky \cOMq \sIM \sooM \sOM   +  \right.\right. \acap
\nnb & + \left.\left. \left(3 \eeM  + 2 \right) \kz  \left(20 \cOOM \sIMq - 3 \right) \right) + \right.\acap
\nnb & +\left.  \left(5 \eeM\cooM   + 6 \eeM  + 4 \right) \sIM \left(3 \ky \cOM - \right.\right.\acap
& - \left.\left. 15 \ky \cOOOM - \kx \sOM + 15 \kx \sOOOM\right)}, \acap
\ang{{\mathrm{K}}_{33}^{\rm (GM)}}_{\PbM} \nnb \lb{KLT33} & = -\rp{3GJ^{'}\nkM}{16 c^2{\aM}^3\ton{1-\eeM}^3}\grf{
-20\eeM \sIIM \sOOM \left(\kx \cOM + \ky \sOM\right)  + \right.\acap
\nnb & +\left. 5 \eeM\cOOM  \left(12 \kz \cIM \sIMq + \ky \cOM \left(3 \sIIIM - \sIM\right) + \right.\right.\acap
\nnb & +\left.\left. \kx \left(\sIM - 3 \sIIIM\right) \sOM \right)  + 2  \left(3 \eeM + 2 \right) \left(3 \kz \cIM + 5 \kz \cIIIM - \right.\right.\acap
&-\left.\left.\left(\sIM + 5 \sIIIM\right) \left(\ky \cOM - \kx \sOM\right)\right)
}, \acap
\ang{{\mathrm{K}}_{12}^{\rm (GM)}}_{\PbM} \nnb \lb{KLT12} & = -\rp{3GJ^{'}\nkM}{64 c^2{\aM}^3\ton{1-\eeM}^3}\grf{
-80\eeM \kz \cIIM \cOOM \sooM  +\right.\acap
\nnb & +\left.  20 \eeM\sIIM \sooM \left(\ky \cOM + 3 \ky \cOOOM + \kx \left(\sOM - 3 \sOOOM\right)\right)   - \right.\acap
\nnb & -\left.  10 \kz \cIIIM  \left(3\eeM \cooM   - 6 \eeM  - 4 \right) \sOOM - \right.\acap
\nnb &-\left. 10 \kz \cIM  \left(5 \eeM\cooM   + 6 \eeM  + 4 \right) \sOOM + \right.\acap
\nnb & +\left. 10  \left(3 \eeM\cooM   - 6 \eeM  - 4 \right) \sIIIM \left(\ky \cOM - \kx \sOM\right) \sOOM + \right.\acap
\nnb &+\left. \left(5 \eeM\cooM   + 6 \eeM  + 4 \right) \sIM \left(\kx \cOM + 15 \kx \cOOOM - \right.\right.\acap
& -\left.\left. \ky \sOM + 15 \ky \sOOOM\right)}, \acap
\ang{{\mathrm{K}}_{13}^{\rm (GM)}}_{\PbM} \nnb \lb{KLT13} & = \rp{3GJ^{'}\nkM}{32 c^2{\aM}^3\ton{1-\eeM}^3}\grf{
-10 \eeM\kz \cooM \left(\sIM - 3 \sIIIM\right) \sOM  + \right.\acap
\nnb & +\left. 40\eeM \sooM \left(\kz \cOM \sIIM + \cIIM \left(\ky \cOOM - \kx \sOOM\right)\right)  -\right.\acap
\nnb &-\left.  4  \left(3 \eeM + 2 \right) \kz \left(\sIM + 5 \sIIIM\right) \sOM + \right.\acap
\nnb & +\left. 5 \eeM\cIIIM  \left(3 \cooM \left(\kx \cOOM + \ky \sOOM\right) + 2  \left(3\eeM + 2 \right) \kx \right) + \right.\acap
\nnb & +\left. \eeM\cIM  \left(5 \cooM  \left(12 \kx \sIMq + 5 \kx \cOOM +5 \ky \sOOM \right)  + \right.\right.\acap
 & +\left.\left. 2  \left(3 \eeM + 2 \right)  \left(20 \left(\kx \cOOM + \ky \sOOM\right)\sIMq + 3 \kx \right) \right)
}, \acap
\ang{{\mathrm{K}}_{23}^{\rm (GM)}}_{\PbM} \nnb \lb{KLT23} & = -\rp{3GJ^{'}\nkM}{64 c^2{\aM}^3\ton{1-\eeM}^3}\grf{
10 \eeM\cooM  \left(-12 \ky \cIM \sIMq + \right.\right.\acap
\nnb & +\left.\left. \ky \left(5 \cIM + 3 \cIIIM\right) \cOOM - 2 \kz \cOM \left(\sIM - 3 \sIIIM\right) - \right.\right.\acap
\nnb &-\left.\left. \kx \left(5 \cIM + 3 \cIIIM\right) \sOOM \right) - 80 \eeM\sooM \left(\kz \sIIM \sOM + \right.\right.\acap
\nnb &+\left.\left. \cIIM \left(\kx \cOOM + \ky \sOOM\right)\right)  + 4  \left(3 \eeM + 2 \right)  \left(-5 \ky \cIIIM - \right.\right.\acap
\nnb & -\left.\left. 2 \kz \cOM \left(\sIM + 5 \sIIIM\right) + \right.\right.\acap
& + \left.\left. \cIM  \left(20\sIMq \left(\ky \cOOM - \kx \sOOM\right) - 3 \ky \right) \right)
}.
\end{align}
%
Note that \rfr{KLT11}-\rfr{KLT23} have  a general validity since they are restricted neither to any specific spatial orientation of the spin axis of $M^{'}$ nor to circular and/or equatorial orbits of $M$ about $M^{'}$.

It should be remarked that indirect, mixed effects of order $\mathcal{O}\ton{c^{-2}}$ arise, in principle, also from the Newtonian tidal matrix of \rfr{KN} when the post-Newtonian effects of the field of $M^{'}$ onto  the  orbital motion of $M$ are taken into account \cite{1989PhRvD..39.2825M}. The same holds also with $J_2^{'}$, accounting for possible deviations of $M^{'}$ from spherical symmetry at the Newtonian level itself. The calculation of such further effects, whose size may be comparable with that of the direct ones, is beyond the scope of this paper.
\section{Possible scenarios of interest for empirical tests}\lb{expe}
Forthcoming space-based missions to astronomical bodies orbiting large primaries such as the Sun and Jupiter, in conjunction with expected progresses in interplanetary tracking techniques \cite{2007IJMPD..16.2117I, 2009AcAau..65..666I, 2014AcAau..94..699I}, may, in principle,  represent an opportunity to put on the test the post-Newtonian tidal effects calculated in the previous Sections.

Let us consider the forthcoming BepiColombo\footnote{See also http://sci.esa.int/bepicolombo/ on the Internet.} \cite{2010P&SS...58....2B} and  JUICE\footnote{See also http://sci.esa.int/juice/ on the Internet.} \cite{2013P&SS...78....1G} missions targeted to Mercury and the Jovian natural satellite Ganymede, respectively; in the following, the symbol $\iota$ will be adopted for the inclinations of the orbital planes of
such spacecrafts to the equators of the orbited bodies. A probe named Mercury Planetary Orbiter\footnote{See http://sci.esa.int/bepicolombo/48872-spacecraft/ on the Internet.} (MPO) \cite{2007PhRvD..75b2001A} is planned to be released in a polar ($\iota_{\rm MPO}=90$ deg), elliptical orbit ($e_{\rm MPO}=0.16 $) around Mercury with an orbital period of approximately $2.3$ h ($a_{\rm MPO} = 3,394$ km). The nominal science duration is one year, with a possible extension of another year. Importantly, orbital maneuvers to change the attitude of the spacecraft are scheduled every about 44 d, so that long smooth orbital arcs should be available.
The JUICE mission \cite{2013P&SS...78....1G} to the Jovian system will culminate in a dedicated orbital tour around Ganymede which should encompass a 30 d science phase during which JUICE will orbit\footnote{See http://sci.esa.int/juice/50074-scenario-operations/ on the Internet.} the satellite in a polar, circular low path with an altitude as little as $h = 200$ km.
Table \ref{Tab:01} summarizes the characteristic frequencies of both the scenarios considered, showing that our results of the previous Sections are applicable to them.
\begin{table}[!t]
\textbf{\refstepcounter{table}\label{Tab:01} Table \arabic{table}.}{ Some characteristic orbital frequencies, in s$^{-1}$, of the Sun-Mercury-Mercury Planetary Orbiter (MPO) and of the Jupiter-Ganymede-Jupiter Ganymede Orbiter (JGO) \cite{2013P&SS...78....1G} systems. For JGO, we considered the planned 30 days phase of the Ganymede tour to be spent in a low altitude (200 km) circular orbit (http://sci.esa.int/juice/50074-scenario-operations/). For the sake of simplicity, the Newtonian and the post-Newtonian precessions due to the oblateness $J_2$ and to the angular momentum $J$ of the primaries were calculated in equatorial coordinate systems. Indeed, while both MPO and JGO will move along polar trajectories   at $\iota=90$ deg to the equators of their primaries, the orbits of Mercury and Ganymede lie  almost in the equatorial planes of the Sun ($\iota_{\mercury}=3.38$ deg) and of Jupiter ($\iota_{\rm Gan}=0.20$ deg), respectively. For Ganymede, the value $J_2^{\rm Gan} =  1.27\times 10^{-4}$ \cite{1996Natur.384..541A} was adopted, while its angular momentum $J^{\rm Gan} = 3\times 10^{30}$ kg m$^2$ s$^{-1}$ was inferred from the values of its mass, equatorial radius and normalized polar moment of inertia \cite{1996Natur.384..541A}. For Mercury, we assumed $J_2^{\mercury}=1.92\times 10^{-5}$ \cite{2010Icar..209...88S}, while its angular momentum $J^{\mercury} =8.4\times 10^{29} $ kg m$^2$ s$^{-1}$ was obtained from the latest determinations of its equatorial radius \cite{2014NatGe...7..301B} and normalized polar moment of inertia \cite{2012JGRE..117.0L09M}. For the angular momentum and the oblateness of the Sun and of Jupiter, we assumed $J^{\odot}=1.90\times 10^{41}$ kg m$^2$ s$^{-1}$ \cite{1998MNRAS.297L..76P}, $J_2^{\odot}=2.1\times 10^{-7}$ \cite{2014IPNPR.196C...1F}, and $J^{\jupiter} = 6.9\times 10^{38}$ kg m$^2$ s$^{-1}$ \cite{2003AJ....126.2687S}, $J_2^{\jupiter} = 1.469\times 10^{-2}$ \cite{Jac2003}, respectively. }
\processtable{ }
{\begin{tabular}{llll}
\toprule
Frequency (Sun-Mercury-MPO) & Value (s$^{-1}$) &  Frequency (Jupiter-Ganymede-JGO) & Value (s$^{-1}$)\\
\midrule
$\dot\omega_{\mercury}^{\left(\rm GM\right)}$ & $1\times 10^{-18}$ & $\dot\Om_{\rm Gan}^{\left(\rm GM\right)}$ & $8\times 10^{-16}$\\ \\
$\dot\omega_{\mercury}^{\left(J_2^{\odot}\right)}$ & $8\times 10^{-17}$ & $\dot\varpi_{\rm Gan}^{\left(J_2^{\jupiter}\right)}$ & $3\times 10^{-9}$\\ \\
$\dot\omega_{\mercury}^{\left(\rm GE\right)}$ & $7\times 10^{-14}$ & $\dot\omega_{\rm Gan}^{\left(\rm GE\right)}$ & $4\times 10^{-14}$ \\ \\
$n_{\mercury}$ & $8\times 10^{-7}$ & $n_{\rm Gan}$ & $1\times 10^{-5}$ \\ \\
$\dot\Om_{\rm MPO}^{\left(\rm GM\right)}$ & $3\times 10^{-17}$ & $\dot\Om_{\rm JGO}^{\left(\rm GM\right)}$ & $2\times 10^{-16}$\\ \\
$\dot\omega_{\rm MPO}^{\left(\rm GE\right)}$ & $2\times 10^{-13}$  & $\dot\omega_{\rm JGO}^{\left(\rm GE\right)}$ & $8\times 10^{-14}$\\ \\
$\left|\dot\omega_{\rm MPO}^{\left(J_2^{\mercury}\right)}\right|$ & $6\times 10^{-9}$ & $\left|\dot\omega_{\rm JGO}^{\left(J_2^{\rm Gan}\right)}\right|$ & $5\times 10^{-8}$\\ \\
$n_{\rm MPO}$ & $7\times 10^{-4}$ & $n_{\rm JGO}$ & $6\times 10^{-4}$ \\
\botrule
\end{tabular}}{}
\end{table}
In Table \ref{Tab:02}, we maximize the values of the post-Newtonian tidal perturbations for both MPO and JGO with respect to  their unknown node $\Om$ and pericenter $\omega$. We do the same also for some of the most important competing Newtonian and post-Newtonian orbital perturbations.
It can be noticed that, while for MPO the gravitoelectric tidal effects are larger than the gravitomagnetic ones, the situation is reversed for JGO because of its scheduled zero eccentricity. As far as the magnitudes of the tidal effects are concerned, they are larger for JGO; its gravitomagnetic tidal precessions reach the $\approx 10^{-1}-10^{-2}$ mas yr$^{-1}$ level.
\begin{table}[!t]
\textbf{\refstepcounter{table}\label{Tab:02} Table \arabic{table}.}{ Maximum nominal values of the direct orbital rates of change $\dot\Psi$ of MPO and JGO, averaged over both $\Pb$ and $\PbM$, induced by the post-Newtonian gravitoelectric and gravitomagnetic tidal  field of $M^{'}$, by its Newtonian tidal field, and by some competing Newtonian and post-Newtonian perturbations due to the deviation from spherical symmetry of the field of $M$.  The values of $\Om_{\rm max}$ and $\omega_{\rm max}$, which are different for each orbital effect considered, are not reported. The units for the precessions are milliarcseconds per year (mas yr$^{-1}$), apart from the eccentricity $e$ whose rate of change is expressed in s$^{-1}$. The  Newtonian $J_2$ and the post-Newtonian $J$ orbital precessions for a generic orientation of the spin axis of $M$ were retrieved from \cite{2011PhRvD..84l4001I}.  The mean equinox and the mean equatorial plane of the Earth at the epoch J2000.0 of the International Celestial Reference Frame (ICRF) were adopted for both $\mathcal{K}$ and ${\mathcal{K}}^{'}$. The orientations of the spin axes  with respect to the ICRF  were retrieved from \cite{2007CeMDA..98..155S}. }
\processtable{ }
{\begin{tabular}{llll}
\toprule
$\dot\Psi$ (MPO) & Value  &  $\dot\Psi$ (JGO) & Value \\
\midrule
$\dot e^{\rm {(tid\ GM)}}$ & $1\times 10^{-21}$ s$^{-1}$ & $\dot e^{\rm {(tid\ GM)}}$ & 0 s$^{-1}$ \\
$\dot I^{\rm {(tid\ GM)}}$ & $4\times 10^{-6}$ mas yr$^{-1}$ & $\dot I^{\rm {(tid\ GM)}}$ & $0.01$ mas yr$^{-1}$ \\
$\dot \Om^{\rm {(tid\ GM)}}$ & $1\times 10^{-5}$ mas yr$^{-1}$ & $\dot \Om^{\rm {(tid\ GM)}}$ & $0.07$ mas yr$^{-1}$ \\
$\dot \omega^{\rm {(tid\ GM)}}$ & $8\times 10^{-5}$ mas yr$^{-1}$ & $\dot \omega^{\rm {(tid\ GM)}}$ & $0.54$ mas yr$^{-1}$ \\
$\dot e^{\rm {(tid\ GE)}}$ & $1\times 10^{-18}$ s$^{-1}$ & $\dot e^{\rm {(tid\ GE)}}$ & $0$ s$^{-1}$ \\
$\dot I^{\rm {(tid\ GE)}}$ & $0.0025$ mas yr$^{-1}$ & $\dot I^{\rm {(tid\ GE)}}$ & $8\times 10^{-7}$ mas yr$^{-1}$ \\
$\dot \Om^{\rm {(tid\ GE)}}$ & $0.0084$ mas yr$^{-1}$ & $\dot \Omega^{\rm {(tid\ GE)}}$ & $2\times 10^{-6}$ mas yr$^{-1}$ \\
$\dot \omega^{\rm {(tid\ GE)}}$ & $0.0458$ mas yr$^{-1}$ & $\dot \omega^{\rm {(tid\ GE)}}$ & $1\times 10^{-5}$ mas yr$^{-1}$ \\
$\dot I^{\rm (GM)}$ & $0.1038$ mas yr$^{-1}$ & $\dot I^{\rm (GM)}$ & $0.59$ mas yr$^{-1}$ \\
$\dot \Om^{\rm (GM)}$ & $0.1907$ mas yr$^{-1}$ & $\dot \Om^{\rm (GM)}$ & $1.24$ mas yr$^{-1}$ \\
$\dot \omega^{\rm (GM)}$ & $0.2075$ mas yr$^{-1}$ & $\dot \omega^{\rm (GM)}$ & $1.18$ mas yr$^{-1}$ \\
$\dot \omega^{\rm (GE)}$ & $1087.78$ mas yr$^{-1}$ & $\dot \omega^{\rm (GE)}$ & $499.6$ mas yr$^{-1}$ \\
$\dot e^{\rm {(tid\ N)}}$ & $2.9\times 10^{-10}$ s$^{-1}$ & $\dot e^{\rm {(tid\ N)}}$ & 0 s$^{-1}$ \\
$\dot I^{\rm {(tid\ N)}}$ & $6.7\times 10^5$ mas yr$^{-1}$ & $\dot I^{\rm {(tid\ N)}}$ & $7\times 10^7$ mas yr$^{-1}$ \\
$\dot \Om^{\rm {(tid\ N)}}$ & $2\times 10^6$ mas yr$^{-1}$ & $\dot \Om^{\rm {(tid\ N)}}$ & $3\times 10^8$ mas yr$^{-1}$ \\
%
%
%
%
%
%
$\dot \omega^{\rm {(tid\ N)}}$ & $1.3\times 10^7$ mas yr$^{-1}$ & $\dot \omega^{\rm {(tid\ N)}}$ & $2\times 10^9$ mas yr$^{-1}$ \\
$\dot I^{ (J_2)}$ & $8\times 10^{6}$ mas yr$^{-1}$ & $\dot I^{ (J_2)}$ & $6\times 10^{7}$ mas yr$^{-1}$ \\
$\dot \Om^{(J_2)}$ & $3\times 10^{7}$ mas yr$^{-1}$ & $\dot \Om^{ (J_2)}$ & $2.7\times 10^{8}$ mas yr$^{-1}$ \\
$\dot \omega^{ (J_2)}$ & $4\times 10^{7}$ mas yr$^{-1}$ & $\dot \omega^{(J_2)}$ & $3.5\times 10^{8}$ mas yr$^{-1}$ \\
\botrule
\end{tabular}}{}
\end{table}
In principle, a rate of $\approx 0.5$ mas yr$^{-1}$, which naively corresponds to a range-rate as little as $\approx 0.05$ mm s$^{-1}$ at the distance of Jupiter from us, might be detectable with the expected improvement down to $0.01$ mm s$^{-1}$ at 60 s integration time  in the Doppler range-rate techniques  from the ASTRA study \cite{2014AcAau..94..699I}. Unfortunately, such figures are too small if compared with those of the competing orbital effects. Suffice it to say that the oblateness of Ganymede is presently known with a relative uncertainty of the order of just $\sigma_{J_2}/J_2 = 2\times 10^{-2}$  \cite{1996Natur.384..541A}. An improvement of $6-7$ orders of magnitude  is beyond the goals of the JUICE mission itself \cite{2013P&SS...78....1G}.
Strictly speaking, such considerations  hold only for the direct post-Newtonian tidal effects calculated in the previous Sections; the total sensitivity budget should account for the indirect, mixed  effects of order $\mathcal{O}\ton{c^{-2}}$ as well.

In principle, a rather unconventional possibility could be the realization of an artificial mini-planetary system to be carried onboard a drag-free spacecraft orbiting, say, the Earth; such an idea was already proposed in the past to accurately measure the Newtonian constant of gravitation $G$ \cite{1987PhLA..120..437N, 1988AJ.....95..576N,  1990ESAJ...14..389N, 1992PhLA..167...29K, 1992PhRvD..46..489S, 1999MeScT..10..514S, 2000CQGra..17.2331S}, and, more recently, to put on the test the MOND theory \cite{2008IJMPD..17..453S}.
The conditions of validity of the present analysis could be fulfilled, e.g., by placing a non-rotating sphere made of tungsten with density $\rho_{\rm W} =19.6$ g cm$^{-3}$ and $M=28\ \mathrm{kg},R=7\ \mathrm{cm}$ inside a drag-free spacecraft orbiting the Earth in some suitably chosen High Earth Orbit (HEO). By assuming, say, $a=10\ \mathrm{cm}$ for the test particle orbiting the tungsten sphere and a geostationary orbit with $\aM=42,164\ \mathrm{km}$ for the spacecraft, it would be possible to obtain
\eqi\rp{\nkM}{\nk}=5\times 10^{-2}.\eqf Moreover, the local dynamics of such a spaceborne  artificial planetary system would be practically free from systematic non-Keplerian gravitational perturbations due to $M$. Indeed, careful manufacturing of the sphere would allow to make its oblateness  negligible; the post-Newtonian gravitoelectric pericenter precession would be completely irrelevant being as little as $\dot\omega\approx 10^{-12}$ mas yr$^{-1}$.
Table \ref{Tab:03} summarizes the nominal maximum values of the precessions of a \virg{planet} orbiting the aforementioned tungsten sphere along a circular path perpendicular to the Earth's equator in a spacecraft following a highly eccentric polar orbit around the Earth.
\begin{table}[!t]
\textbf{\refstepcounter{table}\label{Tab:03} Table \arabic{table}.}{ Maximum nominal values of the direct orbital rates of change $\dot\Psi$ (here, $\varpi\doteq\Om + \omega$ is the longitude of the pericenter) of a member of a spaceborne artificial mini-planetary system orbiting a tungsten sphere of mass $M=28$ kg and radius $R=7$ cm  in a circular orbit with $a=10$ cm, $e=0,\ I=90$ deg averaged over both $\Pb$ and $\PbM$, induced by the post-Newtonian gravitoelectric and gravitomagnetic tidal  field of the Earth and by its Newtonian tidal field.  For the spacecraft hosting it we assumed an highly elliptical geostationary polar orbit characterized by $\aM = 42,164$ km, $\eM=0.7,\ \IM=\oM=90$ deg, $\OM=0$ deg. The values of $\Om_{\rm max}$ and $\omega_{\rm max}$, which are different for each orbital effect considered, are not reported. The units for the precessions are milliarcseconds per year (mas yr$^{-1}$), apart from the eccentricity $e$ whose rate of change is expressed in s$^{-1}$.  The mean equinox and the mean equatorial plane of the Earth at the epoch J2000.0 of the International Celestial Reference Frame (ICRF) were adopted for both $\mathcal{K}$ and ${\mathcal{K}}^{'}$. The orientation of the Earth's spin axis  with respect to the ICRF  was retrieved from \cite{2007CeMDA..98..155S}. }
\processtable{ }
{\begin{tabular}{ll}
\toprule
$\dot\Psi$  & Value  \\
\midrule
$\dot e^{\rm {(tid\ GM)}}$ & $0$ s$^{-1}$  \\
$\dot I^{\rm {(tid\ GM)}}$ & $0$ mas yr$^{-1}$  \\
$\dot \varpi^{\rm {(tid\ GM)}}$ & $3.2$ mas yr$^{-1}$  \\
$\dot e^{\rm {(tid\ GE)}}$ & $0$ s$^{-1}$  \\
$\dot I^{\rm {(tid\ GE)}}$ & $32.8$ mas yr$^{-1}$  \\
$\dot \varpi^{\rm {(tid\ GE)}}$ & $189.6$ mas yr$^{-1}$ \\
$\dot e^{\rm {(tid\ N)}}$ & $0$ s$^{-1}$  \\
$\dot I^{\rm {(tid\ N)}}$ & $2.6\times 10^{10}$ mas yr$^{-1}$  \\
$\dot \varpi^{\rm {(tid\ N)}}$ & $1.5\times 10^{11}$ mas yr$^{-1}$  \\
\botrule
\end{tabular}}{}
\end{table}
In fact, the size of the post-Newtonian tidal effects are not negligible. Nonetheless, the tidal precessions of Newtonian origin would  overwhelm them  since the gravitational parameter $GM^{'}$ of the Earth is currently known with a $2\times 10^{-9}$ relative accuracy \cite{2010ITN....36....1P}, insufficient by $1-2$ orders of magnitude for our purposes.
\section{Overview and conclusions}\lb{concludi}
We looked at the direct long-term orbital rates of change occurring within a local gravitationally bound two-body system as gradiometers to potentially detect post-Newtonian tidal effects due to its slow motion in the external field of a distant third body. We also assumed that the characteristic orbital frequencies of  the internal dynamics of the local binary are quite smaller than the frequency of its orbital motion around the external source. We obtained general analytical expressions valid for arbitrary orbital configurations and for a generic orientation of the spin axis of  the external body. Future work should be devoted to the calculation of the indirect, mixed post-Newtonian effects arising from the interplay between the Newtonian tidal matrix and the post-Newtonian orbital motion of the binary in the external field.

We applied our results to the future BepiColombo and JUICE man-made missions to Mercury and Ganymede, respectively. It turned out that that, although the expected improvements in interplanetary tracking may, perhaps, allow for a detection of the tidal effects we are interested in, especially for JUICE, the impact of several competing orbital effects of Newtonian and post-Newtonian origin,  acting as sources of potential systematic errors, should be carefully considered.

Another possibility which, in principle, may be further pursued is the realization of an artificial mini-planetary system to be carried onboard an Earth-orbiting drag-free spacecraft. If, on the one hand, the post-Newtonian tidal precessions occurring in such a system may be relatively large, amounting to about $1-10^2$ mas yr$^{-1}$, on the other hand, the product of the Earth's mass times the Newtonian gravitational constant is currently known with  insufficient accuracy to allow for an effective subtraction of the competing Newtonian tidal precessions.

\bibliographystyle{frontiersinHLTH&FPHY} 
\bibliography{gradiobib}

\end{document}